\documentclass[graybox]{svmult}

\usepackage{type1cm}        
\usepackage{url}
\usepackage{makeidx}         
\usepackage{graphicx}        
\usepackage{multicol}        
\usepackage[bottom]{footmisc}
\usepackage{todonotes}

\usepackage{newtxtext}       %
\usepackage{newtxmath}       

\usetikzlibrary{shapes, arrows.meta, positioning}    
\usepackage{listings}        
\lstdefinestyle{tree}{
    literate=
    {├}{{\smash{\raisebox{-1ex}{\rule{1pt}{\baselineskip}}}\raisebox{0.5ex}{\rule{1ex}{1pt}}}}1 
    {─}{{\raisebox{0.5ex}{\rule{1.5ex}{1pt}}}}1 
    {│}{{\smash{\raisebox{-1ex}{\rule{1pt}{\baselineskip}}}\raisebox{0.5ex}{\rule{1ex}{0pt}}}}1 
    {└}{{\smash{\raisebox{0.5ex}{\rule{1pt}{\dimexpr\baselineskip-1.5ex}}}\raisebox{0.5ex}{\rule{1ex}{1pt}}}}1 
}

\makeindex             

\begin{document}

\title*{Open Science in Software Engineering}
\titlerunning{Open Science in Software Engineering}

\author{Daniel Mendez, Daniel Graziotin, Stefan Wagner, and Heidi Seibold}
\authorrunning{Daniel M\'endez Fern\'andez et al.} 
\institute{Daniel Mendez \at Technical University of Munich, Germany\\ Blekinge Institute of Technology, Sweden \\ fortiss GmbH, Germany \\
ORCID: 0000-0003-0619-6027 
\email{mendezfe@acm.org}
\and Daniel Graziotin \at University of Stuttgart, Universit{\"a}tsstra{\ss}e 38, 70569 Stuttgart, Germany. \email{daniel.graziotin@iste.uni-stuttgart.de}
\and Stefan Wagner \at University of Stuttgart, University of Stuttgart, Universit{\"a}tsstra{\ss}e 38, 70569 Stuttgart, Germany. \email{stefan.wagner@iste.uni-stuttgart.de}
\and Heidi Seibold \at Ludwig-Maximilians-University Munich, Marchioninistra{\ss}e 15, 81377 Munich, Germany. \email{hseibold@ibe.med.uni-muenchen.de}}

\maketitle

\abstract{Open science describes the movement of making any research artefact available to the public and includes, but is not limited to, open access, open data, and open source. While open science is becoming generally accepted as a norm in other scientific disciplines, in software engineering, we are still struggling in adapting open science to the particularities of our discipline, rendering progress in our scientific community cumbersome. In this chapter, we reflect upon the essentials in open science for software engineering including what open science is, why we should engage in it, and how we should do it. We particularly draw from our experiences made as conference chairs implementing open science initiatives and as researchers actively engaging in open science to critically discuss challenges and pitfalls, and to address more advanced topics such as how and under which conditions to share preprints, what infrastructure and licence model to cover, or how do it within the limitations of different reviewing models, such as double-blind reviewing. Our hope is to help establishing a common ground and to contribute to make open science a norm also in software engineering.}

\section{Introduction}

In a nutshell, open science refers to the movement of making any research artefact available to the public. This ranges from the disclosure of software source code (``open source'') over the actual data itself (``open data'') and the material used to analyse the data (such as analysis scripts, ``open material'') to the manuscripts reporting on the study results (``open access'').\footnote{Open science and open scholarship encompass a wide range of topics and activities, many of which are described by Tennant et al.~\cite{Tennant2019} In this chapter, we concentrate on topics we believe to be in scope of (empirical) software engineering, namely open access, open data, open materials, open source, open peer review, and registered reports}. Disclosing research artefacts increases transparency and, thus, reproducibility and replicability of our scientific process and our results.  Open science is often seen as an important means to move forward as a scientific research community. Open data and open source -- both being major principles under the common banner of open science -- constitute a major hallmark in making empirical studies transparent and understandable to researchers not involved in carrying out those studies. This can be done, for example, by sharing replication packages that capture the raw data and anything necessary for their analysis and interpretation. That way, we increase the reproducibility of our research. This, in turn, strengthens the credibility of the conclusions we draw from the analysed data and it allows others to build their own work upon ours; hence, it strengthens more generally our overall body of knowledge in the research community.

Besides these more ideological views on open science and reasonable arguments in favour of engaging into it as a research community, on which any reader will probably agree, there is much more to it which we need to understand when considering open science in the context of software engineering research. There are, for example, various challenges in data disclosure -- technical ones, ethical and legal ones, but also social ones -- which are different to the standards and views given in other disciplines and which make open science difficult to become the norm in our own field. Consider, for example, the notion of repeatability, replicability, and reproducibility by considering the terminology as introduced by the ACM\footnote{\url{https://www.acm.org/publications/policies/artifact-review-badging}} (verbatim):

\begin{itemize}
    \item \textbf{Repeatability (Same team, same experimental setup):} The measurement can be obtained with stated precision by the same team using the same measurement procedure, the same measuring system, under the same operating conditions, in the same location on multiple trials. For computational experiments, this means that a researcher can reliably repeat her own computation.
    \item \textbf{Replicability (Different team, same experimental setup):}
    The measurement can be obtained with stated precision by a different team using the same measurement procedure, the same measuring system, under the same operating conditions, in the same or a different location on multiple trials. For computational experiments, this means that an independent group can obtain the same result using the author's own artefacts.
    \item \textbf{Reproducibility (Different team, different experimental setup):} The measurement can be obtained with stated precision by a different team, a different measuring system, in a different location on multiple trials. For computational experiments, this means that an independent group can obtain the same result using artefacts which they develop completely independently.
\end{itemize}

As an engineering discipline heavily inspired by the natural sciences, we often make implicit assumptions that our focus is on quantitative and even purely computational studies (e.g. simulations). For these, existing definitions and norms hold as they are and we are able to yield replicability and reproducibility. This situation is, however, not the norm. Most studies in software engineering involve -- in one form or another -- humans. In the end, software is made by human beings for human beings. Human subjects, however, act purely rational in exceptional cases only, if at all~\cite{Lambert06}. This means that every change in an experimental context, even if strictly following the same experimental setup and procedure, will eventually yield different (context-dependent) results. Such studies would then not fit the available definition of reproducibility as used in computational studies, but it is still reasonable to argue that they would be reproducible. Further challenges in software engineering research are that much of our data emerges from sensitive (e.g. industrial) settings and finally the reliance upon qualitative data where the data analysis is less procedural when compared to quantitative data (also imposing significant integrity challenges). All this renders full disclosure often difficult and we often need to anonymise the data to act within legal and ethical constraints that most computational studies do otherwise not have. Those two facets of software engineering research alone show already that we need to adapt open science principles to the particularities of our discipline, same as it is the case in other disciplines.

How can our software engineering community of researchers adopt its own open science movement? We believe that it is a lack of proper understanding about
\begin{itemize}
    \item what open science is (and what it isn't) for software engineering,
    \item why we should all do our best to implement it, whether as editor, chair, or as researcher, and finally 
    \item how we could and should do it
\end{itemize}
that often leads to a general reluctance towards implementing open science. Sometimes, it even leads to a general dismissal of the potential open science has for individual researchers and the community as a whole. All this renders our own open science movement cumbersome.

In this chapter, we cover the essentials in open science for software engineering. In particular, we establish a common ground in our discipline by elaborating on established key terms, principles, and approaches in Sect.~\ref{sec:what} -- all tailored to the particularities of our discipline. We further discuss why we should engage in open science (Sect.~\ref{sec:why}) before discussing practical guidelines to implementing open science in Sect.~\ref{sec:how}. In Sect.~\ref{sec:challenges}, we then end with a discussion of chosen challenges and pitfalls. The latter is based on our shared experiences emerging from open science activities and lessons we learnt so far as authors and as organisers where we implemented first open science initiatives in the empirical software engineering community.

The main target audience consists of software engineering scholars interested in the general notion of open science and those interested in implementing open science in their own research practices. One hope we associate with this chapter is not only to oppose those critical voices still sceptical towards open science, but also to strengthen the voices of those supporting it out of the firm conviction that open science should soon become the norm in software engineering research, too.

\section{What is Open Science?}
\label{sec:what}

Open science is a movement whose aim is to render all artefacts borne out of scientific research activities accessible, without any barriers, to any individual on Earth~\cite{Woelfle2011}. Open science refers also to the scientific part of the broader terms of open scholarship, i.e. ``the process, communication, and re-use of research as practised in any scholarly research discipline, and its inclusion and role within wider society''~\cite{Tennant2019}. Open science itself is an umbrella term that encompasses several facets of openness, for example open access, open data, open source, open government, open notebooks, or open standards~\cite{FOSTER2019}. In the following, we discus those concepts particularly relevant to the (empirical) software engineering research community.

\subsection{Open Access\label{ssec:whatis:open-access}}

Open access is associated with publications, i.e., research articles, technical reports and papers in general. Open access occurs whenever a publication is freely available on the public Internet without any access barrier -- financial, legal or technical ones (including even not to force users to register to systems). It allows individuals to read, download, copy, distribute, print, search or link to the full texts of publications for any lawful purpose~\cite{BOAI2002}. Minor constraints over redistribution and reuse of the publication may still apply and usually take the form of attribution. It is typical with open access publications that the authors retain the copyright of their work, and the act to render the work as open access is enabled through proper licences. The \emph{Creative Commons} licence model is the most widely employed one for open access (see also Sect.~\ref{ssec:whatis:open-data}).

Open access can take several forms according to which version of a publication is made public and at which point of the academic writing process it is made public. If authors make an own produced copy of their work openly available, they perform an act of \emph{self-archiving}. The work is called \emph{preprint} if it reflects a version of their manuscript that has not yet been accepted for publication at a scientific venue. If the content of the own produced work is identical to the content of the accepted publication, it is called \emph{postprint}. The only differences between the \emph{postprint} and the manuscript formally published by a traditional publisher like ACM, IEEE, or Springer is in typesetting differences and the location of the document. The location of pre- and postprints is typically an open repository for pre- and postprints, in contrast to the digital libraries of the publishers. One such example is given in the following while we will go more into detail in Sect.~\ref{sec:how}. 

\begin{question}{Self-archiving via arXiv}
arXiv, pronounced as \emph{archive} and available at \url{https://arXiv.org}, is a repository, born in 1991, of freely accessible preprints and postprints, as well as whitepapers, covering several scientific fields including physics, mathematics, and computer science~\cite{ginsparg2011twenty}. arXiv is free to access, to register to, and to submit to, but it presents two safe guards for publishing. First, authors have to be endorsed by existing members before they are allowed to register in the system. Second, every submission is moderated by volunteers who check for issues such as scope or copyright. arXiv is the de-facto standard repository for mathematics and physics, and with some authors only publishing their work in there, it receives more than 10,000 submissions per month and is, at the time of writing this chapter, hosting approximately 1.5M manuscripts in a distributed archived system of multiple digital libraries all over the world.
\end{question}

The act of self-archiving is also known as \emph{green open access} and it is allowed by the majority of academic publishers with some regulations. 

\begin{question}{Self-archiving options and publishers' regulations}
Different publishers define different regulations with effect to the needs and possibilities of self-archiving, and it is imperative to strictly adhere to these rules. The SHERPA partnership, a partnership of several universities with the original goal of setting up an institutional open access repository, offers with \emph{RoMEO} -- \url{http://www.sherpa.ac.uk/romeo} --  a tool summarising publishers' copyright and archiving policies. RoMEO distinguishes different categories via the following colour codes commonly adopted also in the wider sense:
\begin{itemize}
\item \textbf{White:} Self-archiving not formally allowed
\item \textbf{Yellow:} Authors can archive preprints (i.e. pre-refereeing)
\item \textbf{Blue:} Authors can archive postprints (i.e. final draft post-refereeing) or publisher's version/PDF
\item \textbf{Green:} Authors can archive preprint and postprint or publisher's version
\end{itemize}

\end{question}

Whenever a publisher renders an accepted publication as openly licensed and available without any restriction whatsoever, the artefact becomes open access under the \emph{gold open access model}. This model often follows an author-pays strategy, but there exist also publishers asking for no article processing charges at all. We refer the reader to the work of Graziotin et al.~\cite{Graziotin2014} for more information on open access and its publishing models.

\subsection{Open Data\label{ssec:whatis:open-data}}
Open data is very similar to open access, but it is applied to any data that was produced in the course of research activities, such as the raw data obtained via a controlled experiment. Openness of data can come in various forms and at different degrees; for instance, while an abstract description of a data set (meta data) could be found and accessed online, it could still be the case that access to the full data set would only be granted upon request and only for specific research purposes carefully selected and laid out by the owners of that data set. Here, we point to the FAIR principles\footnote{See also ~\url{https://www.force11.org/group/fairgroup/fairprinciples}} which describe how data should ideally be made open: When data sets are Findable, Accessible, Interoperable, and Reusable, we refer to it as  ``FAIR data''.

In general, open (FAIR) data follows the idea that research data should be freely available to everyone to use and redistribute as they wish, without any restriction whatsoever born out of copyright and licences~\cite{Auer2007}. As with open access, the Creative Commons deeds are commonly employed licences for open data. 

\begin{question}{Creative Commons (CC) copyright licences}
Creative Commons copyright licences (see \url{https://creativecommons.org/licenses/}) constitute a public licence model with the aim to facilitate granting copyright permissions to published work. The two most employed Creative Commons deeds are the Public Domain (CC0, ``No rights reserved'') and the Attribution 4.0 (CC BY 4.0) licence. The former is a licence that implements true public domain, effectively acting as a renounce of any copyright on the artefacts. The latter is an open licence that allows reuse and redistribution of the artefact with the only condition of attributing the original work to the authors.
\end{question}

Besides the frequently used CC licence models introduced above, further ones are possible, too. One example is the Attribution-NonCommercial 4.0 (CC BY NC 4.0), which adds the clause that the original artefact and any derivation of it cannot be used for commercial purposes. While the Public Domain and the CC BY NC licences might seem more suitable for academic work, opting for them can be problematic as we explain in Sect.~\ref{ssec:challenges:right-license}.

\subsection{Open Source}
Open source in open science is nothing different to open source software as it is commonly known by the computer science community. In fact, many argue that the open source software movement served as an inspiration for more openness in various fields going beyond software-related ones (see also the work by Boisseau et al.~\cite{boisseau_omhover_bouchard_2018} providing an elaborate discussion). In any case, several research endeavours in computer science and empirical software engineering, but also other disciplines as well, produce software. One such example is what is often referred to as \emph{research software} (or scientific software), i.e. software products developed with the purpose of analysing (empirical) data, such as Python code. In principle, the software developed can be released as open source software using known licences such as the MIT licence or the GPLv3.

\subsection{Preregistration of Studies}

Preregistration is a useful tool to ensure a certain level of quality of a study design, e.g. by making sure that hypotheses of a confirmatory study were actually pre-defined rather than being defined after having analysed the data to fit the results. Researchers define what their research questions are, why they want to pursue the research, and how exactly they will try to answer their questions. The Open Science Framework is currently one of the most common places to preregister research projects (see \url{https://osf.io/prereg/}). 
Some journals have reported already how preregistration avoids 
\begin{itemize}
    \item publication bias \cite{Dickersin1990},
    \item p-hacking \cite{Head2015extent}, and
    \item HARKing (Hypothesizing after the results are known \cite{Kerr1998harking}).
\end{itemize}

These journals offer the possibility of submitting a \emph{registered report} to their journal.\footnote{For a guide on writing registered reports, we refer the reader to \url{https://osf.io/8mpji/}} Such a report goes through peer review and, provided acceptance, the report is \emph{in principle accepted} (IPA). If the researchers conduct the study as indicated in the registered report, their paper will be published in the journal regardless of the results.

\subsection{Open Science Badges}
For every form of open science, publishers can award \emph{open science Badges}. Badging is a form of promoting open science activities of researchers via a specific badge that publicly recognises their open science engagement. To this end, publishers associate a specific symbol (i.e. a badge) to chosen artefacts to certify that the content is available and accessible in a persistent location.

There exist various forms of badges obeying the particularities of the various available badge systems. Some of them are publisher-specific (such as the ACM badge system\footnote{See \url{https://www.acm.org/publications/policies/artifact-review-badging}}) and some of them are independent, such as the OSF Open Science Badges.

\begin{question}{OSF Open Science Badges}
A wide-spread open science badge system is the one introduced by the Open Science Foundation (OSF, \url{https://osf.io/}) and further promoted by the Center for Open Science (\url{https://cos.io}). This model distinguishes between badges in the following categories:
\begin{itemize}
    \item \textbf{Open Data:} This badge is awarded when shareable data necessary to reproduce a study are made publicly (digitally) available. 
    \item \textbf{Open Materials:} This badge is awarded when making available the materials of the followed research methodology necessary to reproduce or replicate that followed methodology (e.g. analysis scripts).
    \item \textbf{Preregistered:} That badge is awarded when preregistering a study design including the description of the research design and study materials.
\end{itemize}
\end{question}
How to award which badges depends on many (often non-trivial) criteria defined by editors and following a specific reviewing model to check the eligibility to obtain the badges. Although badges are, at the time of writing this chapter, rather rare in software engineering research (such as badges for preregistered studies) and although some systems may still be perceived as difficult to implement (such as the ACM system due to the wide spectrum of often overlapping badges), badges are generally recognised to be a valuable incentive that increases the participation in open science initiatives~\cite{rowhani2017incentives}. Hence, they are being adopted more and more by journals and conferences.

\subsection{Open Peer Review\label{ssec:whatis:open-peer-review}}
Different models of peer review exist and have been experimented with lately~\cite{Tennant2017}. One of these is open peer review, for which there is, however, yet no commonly accepted and clear definition nor an agreed schema as elaborated in a secondary study by Ross-Hellauer~\cite{RossHellauer2017}. Open peer review implementations intend to make the review process as transparent as possible and can feature factors ranging from removing the anonymity of authors and reviewers alike, over making the actual reviews public and allowing for interaction between authors and reviewers, to crowdsourcing reviews and even making manuscripts public before the review phase.

One least common denominator of open peer review focuses on the names of authors and reviewers so that both can see each others' identities. This allows for authors and reviewers to have a direct conversation rather than having to go through third parties for communication purposes (e.g. via handling editors or chairs). In the programming community, this type of review process has long been known in code reviews, but -- despite the advantages recognised in the research community as shown in a recent study on the future of peer review in software engineering~\cite{prechelt2018community} -- it is not yet adopted by our journals and conferences (see also Sect.~\ref{sec:challenges}). One exception is the Journal of Open Source Software.\footnote{For details, see \url{https://joss.readthedocs.io/en/latest/submitting.html#the-review-process}} Another definition focuses on disclosing the reviews -- sometimes with the names of the reviewers. That way, reviewers can be held more accountable, but they can also serve to make the decision for acceptance more transparent to others and the reviewers can also claim the recognition they deserve. There are many fears and hopes around open peer review models, many of which are discussed in an editorial of the European Journal of Neuroscience after having implementing such a model \cite{doi:10.1111/ejn.13762}. One fear (for which, however, there is no evidence yet) is the risk that early career researchers might be more reluctant to provide profound critique if their names are revealed (see also our discussion in Sect.~\ref{sec:SharingPreprints}). A partial implementation of this model where reviewer names and their reviews are made public is followed by the PeerJ Computer Science Journal, which asks the reviewers whether they wish to disclose their name and subsequently to the authors if whether they wish to disclose the peer review history in the published paper.

\section{Why do we need Open Science?}
\label{sec:why}
Open science is becoming more and more accepted in scientific communities to be having many positive effects. These effects range from increased access and citation counts~\cite{eysenbach2006citation} to facilitating technology transfer with the industry and fostering collaborations through open repositories. Academic publishing and knowledge sharing is meant to become more cost-effective -- German university libraries alone are estimated to be spending well beyond 200 million EUR on publication subscriptions fees per year~\cite{schimmer2015disrupting} -- and researchers and practitioners with no publisher subscriptions can freely access and build on the work of others. There are many discussions and controversies centred around publisher subscription models and how institutions (and institutional alliances) should deal with them. In this chapter, we will not even try to address these discussions to the extent they deserve, but provide a broader view on why we do need open science in general.

Imagine the following situation: A conference author submits a manuscript promising to have provided scientific and empirically-informed arguments for considering Go To statements harmful; a statement previously relying on rationalist arguments of software engineering pioneers like Dijkstra~\cite{Dijkstra68} only. As laid out by that author, those arguments emerge from the exploration of industrial source code -- which the author does not share, maybe because of non-disclosure agreements with collaborating companies from which the data emerges, or maybe for other reasons; this statement is not made explicit in the manuscript. They have further analysed the impact of those statements based on in-depth interviews -- which the author does also not share, maybe because of ethical and legal constraints. Imagine further that the reviewers find no obvious methodological flaws in the design which the author describes in great detail for both the content analysis and the interviews. The author is an experienced and recognised authority in the research community and the manuscript is written in an easy-to-follow manner. The reviewers further find the manuscript ``compelling'', ``interesting'', and the results are also ``surprising'' to them given the availability of contrary evidence provided by other authors who previously analysed publicly available software repositories coming to very contrary conclusions~\cite{NRY+15}. Even if the submitting author did not discuss that other publication in detail, a presentation of that work would certainly lead to controversial and interesting discussions; something the reviewers believe to merit presentation at the prestigious conference they review for. So they recommend acceptance and the PC chairs select that publication for inclusion in the program. It is reasonable to believe that many readers of this chapter having served as co-chairs and reviewers for conferences can identify with such a situation.

Now imagine you were a young scholar analysing the effects of software defects and you find this publication. You would certainly find this publication interesting as it could provide a useful ground for follow-up work. Ask yourself -- honestly -- the following questions:

\begin{itemize}
    \item Would you trust the results? If so, based on what? The simple fact that it has been accepted by the prestigious conference? The way the manuscript is generally written? The name of the author or her or his affiliation? Maybe based on the high number of citations that this publication already has? Maybe it is a combination of all factors? Would the picture change if the author would be unknown to you and if the work would have been published at a lower ranked conference?
    \item Would you be able to really comprehend how the study has been carried out? Would you be able to reproduce the conclusions drawn by the author based on the insights provided in the manuscript? Would you be able to replicate the study in your own research environment?
    \item To what extent does that piece of work provide a good theory for your work? Would this theory be robust and reliable (i.e. scientific)? Would you consider it useful?
    \item How would you use the work if you could only access the abstract of the manuscript because it is hidden behind a paywall and because your institution has no subscription? Would you cite the work based on the information in the abstract? Maybe based on the statements found in other papers citing that work?
    \item How would you cite that work and put it in relation to your own research? Would the picture change in dependency to whether the statements in that manuscript support your own arguments or whether it contradicts them?
\end{itemize}

This very example certainly describes a fictitious situation and yet it describes in many ways the de-facto situation of software engineering research. Scientific practices are -- and they need to -- rely on certain safe guards, such as peer review, but they are nevertheless also dictated by social and political mechanisms and many non-trivial, subjective factors in the research communities. These factors very often dictate in one form or the other which submissions eventually make it into the publication landscape and which do not, and which publications are cited and which are not. As a consequence, publication and citation regimes -- although inherently rooted in scepticism -- have also much to do with trust and convictions~\cite{MendezPassoth18}; something which holds for most, if not all, scientific disciplines. Transparency is therefore key to break with scientific theories being grounded in common sense, taken-for-granted knowledge, hopes, convictions, and provisional beliefs.

Software engineering still faces many challenges other scientific disciplines do not face. Our data comprehends qualitative and quantitative data types and the theories we work on often have various disciplinary backgrounds (from mathematics over psychology to sociology). Further, our data very often emerges from highly sensitive environments making a disclosure difficult and in many cases impossible. Even if we can disclose the data, in many cases it has to be anonymised to an extent it becomes difficult to fully comprehend. All this renders building and evaluating empirically grounded theories in our field difficult. Hence, scientific practices often remain rooted in trust rather than being rooted in transparent scientific processes. Yet and as laid out by Mendez and Passoth~\cite{MendezPassoth18}, it is theory building which constitutes a crucial foundation to our avenue towards turning our engineering discipline into a more scientific, evidence-based one, same as it was the case for many other disciplines before. Transparency, credibility, and reproducibility are cornerstones in building and evaluating robust and reliable theories for our still emerging field and open science provides a solid foundation to achieve that goal.

In essence, open science practices in general and data sharing in particular eventually allow us as a community of software engineering researchers and practitioners to effectively make contributions to our body of knowledge based upon shared data sets -- making our empirical studies transparent, comprehensible, and credible -- thus, we move forward as a community. As we argue, not only scientific publishing is essential in knowledge sharing and dissemination~\cite{houghton2010economic}, but it is an essential facet in accumulating knowledge via a variation of studies tackling the same or similar questions and building upon the same or similar settings and data sets -- e.g. as part of replication studies~\cite{GJV10} which are rendered difficult if not impossible without clear open science principles dictating shared values and principle scientific practices. 

Therefore, there is no doubt anymore \emph{whether} open science will become the norm also in software engineering research. Ever more public and private funding bodies are implementing open access and open data policies~\cite{childs2014opening, van2011managing}. Also the research community is in tune with with this movement, as we can observe: editors and conference organisers are already planning for a smooth transition to open data, and reviewers are becoming more and more sceptical towards manuscript submissions which do not disclose their data and, consequently, ask the reviewers for too much credit. 
It remains, however, often still a question of \emph{how} the community should adopt open science practices and how individual researchers should open their research. We discuss this question in more detail in the next section.

\section{How do we do Open Science?}
\label{sec:how}

In the following, we address the question of how to engage in open science. There are many aspects to consider when engaging as a researcher in open science. We believe that these aspects are best introduced along a simple (again, fictitious) scenario introduced next. The goal is to show demonstrate opportunities along an exemplary set of practices and techniques available to engage in open science in a hands-on manner.

\subsection{Exemplary Scenario}
As an exemplary scenario, we consider a research project where we are researchers at European universities collaborating with project partners from other universities in the United States. Those partners are researchers in psychology. Our project aims at conducting a psychometric software engineering study and our overall goal is to collect data involving a large-scale study with human subjects. The research design is done in a joint effort. While our partners are largely responsible for the study execution and the data collection, we are largely responsible for analysing the data and reporting on it. 

To keep the example simple, we focus on the statistical analysis of quantitative data in our study, but also refer the reader to the challenges emerging from the disclosure of qualitative data in Sect.~\ref{sec:challenges}.

\subsection{Overall Data Analysis Process}

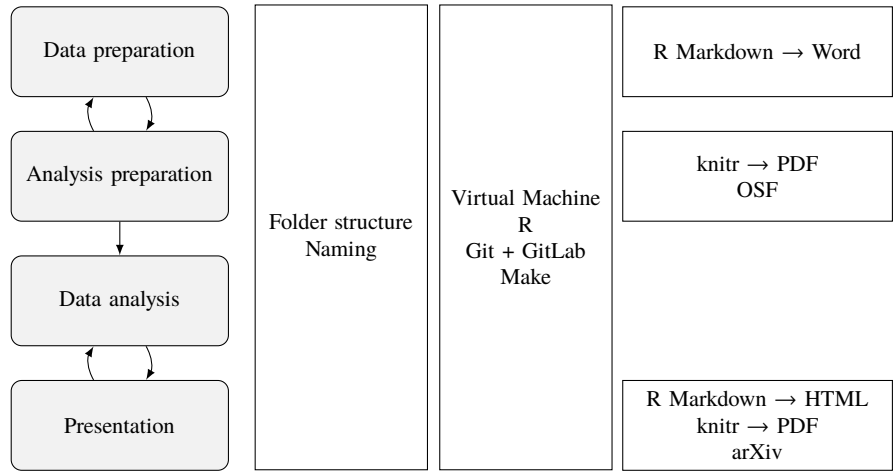
\begin{figure}[htb!]
    \centering
    \tikzstyle{block} = [rectangle, 
    draw, 
    fill=gray!10, 
    text width=9em,
    text centered, 
    rounded corners, 
    minimum height=4em]
\tikzstyle{rec} = [draw, 
    rectangle, 
    text width=10em,
    text centered, 
    node distance=3cm,
    minimum width = 12em, 
    minimum height=4em]
\tikzstyle{line} = [draw, -latex]
    
    \begin{tikzpicture}
        \node [block] (1) {Data preparation};
        \node [block, below= 1.5em of 1] (2) {Analysis preparation};
        \node [block, below= 1.5em of 2] (3) {Data analysis};
        \node [block, below= 1.5em of 3] (4) {Presentation};

        \path [line, bend left] (1) to (2);
        \path [line, bend left] (2) to (1);
        \path [line] (2) to (3);
        \path [line, bend left] (3) to (4);
        \path [line, bend left] (4) to (3);

        \node [draw, 
               rectangle, 
               right= 5em of 4,
               text width=7em, 
               text centered,
               anchor=south, 
               yshift=-2em, 
               minimum height=20.7em] (tb) {Folder structure\\ Naming};
        
        \node [draw, 
               rectangle, 
               right= 0.5em of tb,
               text width=7em, 
               text centered,
               minimum height=20.7em] (ta) {Virtual Machine\\ R\\ Git + GitLab\\ Make};
        
        \node [rec, right= 17.5em of 1] (t1) {R Markdown $\rightarrow$ Word};
        \node [rec, right= 17.5em of 2] (t2) {knitr $\rightarrow$ PDF\\  OSF};
        \node [rec, right= 17.5em of 4] (t4) {R Markdown $\rightarrow$ HTML\\ knitr $\rightarrow$ PDF\\ arXiv};
    \end{tikzpicture}
    \caption{Schema of an exemplary simple project.}
    \label{fig:project}
\end{figure}

Figure~\ref{fig:project} depicts, on the left side, the steps followed in our data analysis with a particular focus on those aspects relevant from an open science perspective. Overall, we first prepare our data and check for any errors, inconsistencies, and missing values, and we discuss these with our partners. At the same time, we start thinking about how to best answer our questions at hand. While we design our analysis procedure, we update the data structure to best fit the analysis plan. Once the analysis plan is finalised, we make it openly available. Ideally, we submit it as a \emph{preregistered study}. This submission includes our study protocol and the material (analysis scripts) as well as a detailed sample description allowing reviewers to judge upon the potential of the study with respect to its theoretical and practical impact. After registering our study and considering the feedback received, only then, we decide to begin with the data analysis.

After discovering no clear patterns in the data, we decide to participate at a workshop where we present our ongoing work based on a previously published short paper describing the overall goal of the study and preliminary results. This work in progress presentation serves the purpose of receiving further feedback from the research community and of getting useful ideas on how to improve our data visualisation techniques. After successfully finishing our data analysis, we finally write up our main publication on the project and disclose our manuscript preprint prior to submitting our manuscript for review to a journal. 

In the following, we walk through that process while focusing on the infrastructure and tools. Our hope is that by presenting the process in such a pragmatic hands-on manner allows to fully reproduce the process as it should typically appear in a research setting.

\subsection{Exemplary Walk-through}

There are various tools to be used to make our project open and reproducible. While we do not claim to be able to present an exhaustive list here, our aim is to give some examples which we use ourselves to make recommendations based on our own experiences. One basic issue to consider first is the folder structure and the naming convention. A good folder structure, in our view, could be like the one in Listing~\ref{lst:prj-structure} as it captures the very essence of our process:

\begin{minipage}{\linewidth}
\begin{lstlisting}[style=tree,caption={Project structure and naming convention for open science},label={lst:prj-structure},language=Bash]
myproject/
    ├── README.md
    ├── Makefile
    ├── data/
    │    ├── clean_data.Rmd
    │    ├── clean_data.docx
    │    ├── data_clean/
    │    │    ├── mydata.csv
    │    │    └── mymetadata.json
    │    └── data_raw/
    │         ├── messy_data1.xlsx
    │         └── messy_data2.csv
    ├── analysis_plan
    │    ├── analysis_plan.Rnw
    │    └── analysis_plan.pdf
    ├── analysis/
    │    ├── analysis.R
    │    └── functions/
    │          └── myfunction.R
    ├── conference_slides.Rmd
    ├── conference_slides.html
    ├── man_references.bib
    ├── manuscript.Rnw
    └── manuscript.pdf
\end{lstlisting}
\end{minipage}

Note that the folder structure clearly defines the different steps shown in Figure~\ref{fig:project} and the folder and file names clearly indicate what each of them contains. 

Regardless of the actual size of the project, the basic rule should be to apply that structure and naming convention concisely and consistently. We experienced it to also be important to keep the original data in a separate folder (\lstinline{data_raw/} in Listing~\ref{lst:prj-structure}) and to not manipulate the raw data files but to create new data files in a separate folder for the data cleaning and analysis (\lstinline{data_clean/} in Listing~\ref{lst:prj-structure}). In combination with a script which cleans the data (\lstinline{clean_data.Rmd} in Listing~\ref{lst:prj-structure}), this makes the data cleaning process reproducible to others.


%
To keep the working environment stable in terms of software versions, we decide to use a \emph{virtual machine} for this project. An alternative option could also be a container (Docker, Singularity, etc.). For the data cleaning and the analysis, we decide to use \emph{R}~\cite{rstats}, an open source software environment for statistical computing. An alternative to that could be to use Python. R scripts (e.g.~\lstinline{analysis.R} in Listing~\ref{lst:prj-structure}) are text files that can be executed in the R console. In contrast to click-and-point programs (e.g.~SPSS when used without syntax) or programs producing binary files (e.g.~Excel), R, same as Python, allows for a reproducible workflow which can be easily version controlled. 

For version control, in our project, we decide to use \emph{Git}~\cite{chacon2014pro} in combination with the Git-repository hosting service \emph{GitLab} (\url{https://gitlab.com}). That version control system allows us and our collaborating partners to trace the versions of all produced text documents in an organised fashion. In combination with the hosting service GitLab, these versions remain available online to all involved in our project. 
For automating our workflow, we use Make~\cite{stallman2001gnu}. To this end, and we keep referring to Listing~\ref{lst:prj-structure}, we store a \lstinline{Makefile} in our main project folder which contains the information on how different files depend on each other, for example that \lstinline{data/clean_data.Rmd} depends on \lstinline{data/data_raw/messy_data1.xlsx} and \lstinline{data/data_raw/messy_data2.csv} and produces \lstinline{data/data_clean/mydata.csv}, \lstinline{data/data_clean/mymetadata.json}, and \lstinline{data/clean_data.docx}. Our \lstinline{Makefile} also documents how the outputs can be produced (via bash commands). 

Next to using R for our project, we use \emph{R Markdown}~\cite{xie2018rmarkdown} and \emph{knitr}~\cite{xie2015dynamic}. Both allow users to combine R code chunks with explanatory text snippets and, thus, allowing for literate programming~\cite{knuth1984literate}. Our text is formatted with Markdown (R Markdown) and LaTeX (knitr). As our partners rely on MS Word, we regularly convert our R Markdown documents to Word documents for constant feedback by commenting directly in those documents. This simplifies the communication about the constant data checking and cleaning process. For an intermediate project report and later for the manuscript writing, we use knitr as it gives us more formatting options. 

Our analysis plan is written with knitr and we upload the PDF to the open science Framework (OSF,~\url{https://osf.io}). This allows us to use the analysis plan for preregistration of the work we aim to do. Preregistration allows to reduce biases in the process of the data analysis (see also \url{https://osf.io/prereg}). We create the slides for the conference again using R Markdown which can produce high quality HTML slides. The manuscript is written using knitr and we make it available as open access on the preprint server arXiv (\url{https://arXiv.org}). To check whether preprint sharing is within the legal constraints of the publisher of the conference, we check for it using the search engine SHERPA RoMEO (\url{http://sherpa.mimas.ac.uk/romeo}).


As we see that the publisher follows a yellow open access model allowing to disclose the preprints but not the postprints, we choose to upload our preprint only. After that submission, we directly submit our manuscript to a peer-reviewed journal. Upon acceptance of the manuscript by that journal, we update our preprint with the DOI provided by the publisher, but do not submit the postprint, i.e, the post production version of the manuscript to comply with the copyright agreement. This preprint version is also the one we distribute among the community, e.g. via social media.

Since all root documents are text files (except for \lstinline{data/data_raw/messy_data1.xlsx}) we can further put them under version control with Git. Through GitLab, we can make them easily accessible to others. This way, our project folder \lstinline{myproject/} can be seen as a replication package. Prior to disclosure, however, we check for parts in our data that need anonymisation to comply with the European data protection regulations (GDPR) as well as with the approval notification of the Institutional Review Board of our partners in the U.S.. We remove any data that might allow to trace observations back to individuals participating in the study.

For our work to be reproducible in a long-term manner, we need to further document the versions of the software used. The virtual machine does that for us, but is not very portable. The option we follow is to use the version management system \emph{packrat} in R~\cite{ushey2018packrat}. 

We notice that our partners are very reluctant to share the data because of its sensitivity and because they fear misuse (e.g. when taken out of its context), thus, we would not be able to follow the FAIR principles (Sect.~\ref{ssec:whatis:open-data}) as anticipated. It is, however, possible for us to convince our project partners to disclose the data when implementing some safeguards. To this end, we decide to disclose our data using the service platform Zenodo (\url{https://zenodo.org}) while choosing \emph{Restricted Access}. Other researchers interested in accessing the data can first read the extensive meta data describing the content of the data and how it was produced. If they believe that the data would fit their scope of interest, they can apply for access and our previously established \emph{Data Use and Access Committee (DUAC)}, formed by us data owners and a member of the responsible ethics committee, so that we can decide whether to grant access to the data or not.

That very example, we hope, illustrates an open science-conform study analysis and reporting producing all artefacts relevant to an open science format adoptable to software engineering and including the disclosure of:
\begin{enumerate}
    \item A study protocol submission and review prior to publication (preregistered study)
    \item The replication package including all analysed data (open data) and all files, scripts, and codebooks necessary to comprehend the study (open materials)
    \item A preprint (yellow open access)
\end{enumerate}
Needless to say, the example is a simplified one neglecting some challenges we typically encounter in practice. In the following, we discuss those challenges in more detail.

\section{Challenges, Pitfalls, and Guidelines}
\label{sec:challenges}

In the following, we discuss typical challenges and pitfalls in open science from the perspective of researchers engaging in open science. To this end, we draw from our experiences covering both the roles of researchers and the roles of organisers (handling editors and conference and workshop organisers).

\subsection{General Issues}

The major challenge that keeps researchers from following all the open science practices described above is probably the difficulty and effort required when making everything openly available. All the practices constitute additional steps that researchers have to do in addition to the non-open research process. They might be motivated to do these additional steps to support the scientific process and higher visibility of open publications. Yet, this motivation has limits. Therefore, the ease of doing open science practices is essential.

In our experience, the difficulty of being open has reduced dramatically over the years. It is easy and cost-free to handle a research project on GitHub or OSF, to permanently publish data on Zenodo or figshare, and to provide preprints on services like arXiv. Some difficulty lies in the details, such as the LaTeX requirements of arXiv, but nowadays we mostly work with modern web applications that behave as one would expect.

Another challenge that might keep researchers from employing openness in their research is the area of conflict between anonymity and confidentiality on the one side and openness on the other. In open science, we ideally would like to make everything open that helps others to understand, verify, and build on our work. When we work with companies, however, they have an understandable interest to protect their intellectual property and reputation, often reflected in signed non-disclosure agreements. Therefore, we have to reduce the data that we can make open or anonymise the data that we have. This is, again, additional effort and a risk that we accidentally make something open that should be confidential.

Similarly, when our studies involve humans, they have an interest in protecting their private data. With the EU GDPR, we now also have a strong legal basis for that. Hence, again, we have the risk to violate corresponding laws. In both cases, companies and individual humans, it is therefore imperative to publish any potentially sensitive data only with the explicit consent of the study participants. Only they themselves can decide what is sensitive and critical for them. In principle, this holds for any kind of publication and, hence, only needs to be extended to ask for consent for publishing the data as well. Anonymising company names is often enough. For anonymising sensitive data of study participants, there are also established techniques (see, e.g., \cite{doi:10.1177/1468794114550439}).

The challenge of anonymity also plays into the third more general issue we would like to mention: Often, openness is merely an afterthought. After we have done all the work, we provide a preprint and make the data available. Ideally, however, the whole process should be open, for example by using OSF or GitHub for all the documents, data, and analysis scripts. In terms of anonymity, this is difficult, as we cannot make everything open and often need a shadow repository with the original raw data. The raw data needs then to be carefully filtered when stored in an open repository. Yet, keeping everything open has the advantage that there is no way of manipulation during the analysis and publication phases of the research. We cannot make the hypothesis fit the data in hindsight because we documented the hypothesis before we did the analysis.

\subsection{Sharing Preprints}
\label{sec:SharingPreprints}

For preprints, we need to consider where we want to publish the paper later on. Upon acceptance of our manuscript, we can also post a postprint. This is rarely a problem when we already have a preprint that is simply updated. Otherwise, there might be publisher-specific embargo periods that need to be adhered to. 

\begin{question}{Self-archiving options for Software Engineering}
In principle, different publishers have different criteria about what they allow at all and what licences to choose. One helpful overview of the different self-archiving options in tune with the regulations of the major publishers in Software Engineering is, as we believe, provided by Arie van Deursen~\cite{VanDeursen2016}.
\end{question}

One challenge we would like to highlight in context of preprint sharing emerges from the trend in software engineering to push for double-blind reviewing models by also anonymising not only reviewers' identigies but also ones of the authors. While the higher goal to reduce potential biases is laudable, it complicated open science practices considerably. Conferences are increasingly adopting a double blind model of peer review, which does not easily allow preprints to be made available because it might allow the reviewers to find out who the authors are. It has been our effort to start a trend in conferences to allow self-archiving preprints and instruct peer reviewers to not actively look for the papers under review online, but it remains nevertheless a challenge. The picture would change if open peer review would be implemented in a code review style (as discussed in Sect.~\ref{ssec:whatis:open-peer-review}). However, the downside and fear of many researchers is that open peer review will put a lot of pressure on researchers, especially early career researchers: Both as authors -- the reviewers will know who made potential mistakes -- and as reviewers -- the authors will know who proposed the changes or even who recommended rejection of the paper.


\subsection{Choosing an Appropriate Licences \label{ssec:challenges:right-license}}

A common pitfall while starting to use open science practices is to assign unsuitable licences. arXiv, for example, allows to select an ad-hoc non-exclusive licence (to arXiv). Granting this minimal licence is compatible with any relevant venue a researcher might want to submit to. Hence, it keeps all options open even if the paper is rejected at the initially planned venue. Adding a Creative Commons licence could reduce this flexibility considerably. In fact, arXiv itself allows to choose from various Creative Commons licences (CC BY, CC BY-SA, CC BY-NC-SA) as well as the CC0 dedication (i.e., public domain)~\cite{arXivLicense2019}. 

Many argue that CC0 is preferable because it frees people from dealing with all attributions. However, in the scientific context, attributing the source and authors of all artefacts that are used is good practice independent of the licence used. PeerJ PrePrints, for instance, enforces the CC BY licence exclusively~\cite{PeerJLicense2019}. This licence is also recommendable for postprints, provided postprint sharing is compatible with the publisher copyright agreement, as it ensures that the researchers are given credit while giving others the largest amount of freedom to share and reuse the manuscript.

In principle, choosing the proper licence is a non-trivial but important task, because certain licences for preprints might cause incompatibility issues further down a publishing chain. Certain licences, including some Creative Commons ones, prevent the work to be used in commercial settings (the -NC part of the CC) or require the redistribution of derivative works using the same licence (the -SA part of the CC). Traditional publishers are, most of the times, commercial entities that require either a full copyright transfer or exclusive rights to distribute the work in a more restricted way, i.e. selling access to papers through paywalls. Non-commercial and share-alike CC licences are, thus, in most of the cases incompatible with traditional publishing models. 

Even the more liberal CC BY licence, which only requires attribution and does not enforce a share-alike clause, might pose issues with traditional publishing as it is non-revocable and allows commercial use by anyone (i.e. non-exclusive to the publisher). The CC0 dedication has also caused issue with traditional publishing in the past~\cite{Russel2011}. The default licence by arXiv is a non-exclusive licence to distribute \cite{ArxivLicenseDistribute2019}, and, virtually, does solely allow arXiv to distribute and display a document (meaning that, theoretically, we are not allowed to do anything at all with arXiv submissions but reading them). This licence is perhaps the most restrictive one among the free licences, making it compatible with traditional publishing (if the copyright transfer conditions allow for it, see Sect.~\ref{ssec:whatis:open-access}). 

We can provide two recommendations. arXiv default non-exclusive licence to distribute should be used when there is certainty to publish a paper with a traditional publisher. A CC-BY licence should be used when there is certainty to publish a paper with a gold open access journal. We do not recommend licensing any preprint, postprint, or dataset using a non-commercial clause (-NC). While counter-intuitive at first sight (we wish for our work to stay free, after all), a non-commercial clause prevents the work to be used by commercial entities. The term \emph{commercial} is, from a legal perspective much broader than it might appear at first; it might affect a large spectrum of people and entities including a simple blog if the website uses an advertisement system. There exist open companies that were born from commercial entities and that are therefore not non-profit (e.g., figshare and PeerJ), and these would not be allowed to make any use of material licensed with the -NC clause. Some of the work might include data mining of papers and datasets and aggregating results, which might still be very useful for the advancement of knowledge. For more information on these legal aspects, we direct the reader to a joint group of copyright experts and Wikimedia~\cite{Wikimedia2012}.

\subsection{Sharing Data and Materials}

A common pitfall in publishing open data and open materials, e.g. as part of replication packages, is to use a personal or institutional website for quickly and easily making them available. It gives one a unique ID in the form of a URL. Yet, a challenge is that we cannot ensure that the URL stays valid and that the content stays on the website. As it has been empirically demonstrated, web pages disappear continuosly~\cite{Koehler2002,Koehler2003}. Therefore, repositories such as Zenodo or figshare, providing a DOI and ensuring permanent archival, are much preferable.

There are small differences between the repositories, but both are recommendable. figshare is commercial but free to use, and its usability seems more polished than at Zenodo. Furthermore, figshare participates in data preservation mechanisms while Zenodo does not. The permanency of Zenodo is ensured, because it is financed by the European Union and run by CERN. 

Similarly as with preprint sharing in the context of double-blind reviewing models, the availability of open data and material would also reveal the authors' identity and, hence, is rendered complicated. While there is no easy solution to the problem of sharing preprints when following a double-blind reviewing model, open data repositories allow researchers now to publish data anonymously for review, thus, being compliant to restrictions imposed by such reviewing model. The authors of the data can then be made public after the paper is accepted. A set of instructions on how to share and archive open data and keep it compatible with double-blind review are presented by Graziotin~\cite{Graziotin2019}.

\subsection{Preparing Qualitative Data}

Achieving replicability and reproducibility of qualitative studies is particularly challenging and many might argue that it is not possible at all (see also the introductory discussion). This renders, however, the disclosure of qualitative data not less important than the disclosure of quantitative data. Even if we cannot support reproducibility of qualitative studies in the nearer sense (if interpreting those terms literally), we can at least achieve transparency of the research and support researchers not involved in the study in understanding how the researchers carrying out the study have drawn their conclusions.

Qualitative data is usually the most difficult to prepare for disclosure in a replication package, because it is most personal and most difficult to anonymise within legal and ethical constraints. A number is more abstract (and easier to open) than spoken words spoken (and transcribed) by individuals, e.g., during an interview. Ideally, we anonymise also qualitative data\footnote{By anomymisation of qualitative data we refer to the removal of any information that allows to reveal the individuals' identities and / or otherwise sensitive not directly related to the study.} and publish it with the explicit consent of the participants. It is important to be open about it upfront to understand whether the participants will agree. Especially for qualitative data, it might often not be the case that we get the consent. Then, it is even more important that at least the analysis material is shared. This is typically easier to share and may include a study protocol as well as the coding schema and coding rules used when coding qualitative data (e.g. as part of a Grounded Theory study). That way, reviewers and other researchers can at least check the trustworthiness of the analysis process and understand how the authors have drawn their conclusions.

\section{Conclusion}
\label{sec:conclusion}
Open science describes the movement to render all artefacts born out of scientific research activities accessible. Openness in our research processes is important to move forward in building reliable and robust theories, thus, turning our discipline into a more scientific one. As outlined in this chapter, we still face, however, various challenges other disciplines do not face. Despite those challenges of adapting open science to the software engineering context, we can still see that our research community is making great progress in that direction. We have ourselves either accompanied or fully implemented efforts to help the community opening up their research artefacts. 

In the course of our endeavour, we have noticed very well that introducing open science into a research community is a difficult and sensitive task, because open science is still often confronted with prejudice, but also because many authors, despite their willingness to conform to such policies, do not often know how exactly to follow such an initiative; that is to say, it is often difficult to see what we should do and what we can do (also considering ethical and legal constraints). 

This is also the reason why we, as organisers, are often constraint by a general reluctance of implementing mandatory open science principles (e.g. via open data policies), thus, rendering the transition to more openness in our discipline rugged. However, the implementations of open science policies in recent editions of conferences and journals -- even if non-mandatory ones were authors could participate on a voluntary basis with the support of dedicated open science chairs --  nevertheless showed high participation ratios with more than 50\% of the authors disclosing their data. Such a support by the community and the positive feedback, e.g. in Town Hall meetings, strengthen our confidence in that the research community is showing more and more awareness of the importance of open science and that open science will eventually become the norm.

One hope we associate with our ongoing efforts in implementing open science initiatives in software engineering venues is to send strong signals into the research community and to gradually increase the awareness of participating researchers to move further in that direction. 

Arguably, we are still confronted with various challenges, such as:
\begin{itemize}
    \item How to implement a uniform and transparent guideline to review disclosed artefacts covering all possible variations in the different types of study (e.g. quantitative and qualitative ones)?
    \item How to implement preregistered studies (which we consider especially important to tackle the problems of publication bias or p-hacking) in tune with the reviewing processes of our existing journals and conferences and how to re-define existing roles and responsibilities?
    \item How to properly reward authors with a clear and easy to understand (and to use) badge system which recognises the differences in the various study types and the difficulties in opening up sensitive, e.g. industrial, data?
    \item How to implement open peer reviews? We can nowadays observe a significant turn in the existing single-blinded reviewing regime, which we applaud, but instead of opening up reviews as well, the current trend is towards even more closeness via double-blind reviewing models, thus, rendering other open science activities difficult, too.
\end{itemize}

We are still convinced that it is not anymore a question whether open science will become the norm also for the software engineering research community, but we recognise that there is still a long way to go, also because we still need to increase the awareness for what open science is, why it is so important, and how to properly adopt such principles to software engineering. 

The chapter at hands is intended to address these questions and to contribute to the movement. Our hope is to further encourage all members of our research community in joining us in this important endeavour of actively shaping an open science agenda for the software engineering community.

\begin{acknowledgement}
We want to thank all members of the empirical software engineering research community actively supporting the open science movement and its adoption to the software engineering community. Just to name a few: Robert Feldt and Tom Zimmermann, editors in chief of the Empirical Software Engineering Journal, are committed to support the implementation of a new Reproducibility \& Open Science initiative\footnote{See also \url{https://github.com/emsejournal/openscience}} -- the first one to implement an open data initiative following a holistic process including a badge system. The steering committee of the International Workshop on Cooperative and Human Aspects of Software Engineering (CHASE) supported the implementation of a open science initiative from 2016 on. Markku Oivo, general chair of the International Symposium on Empirical Software Engineering and Measurement (ESEM) 2018, has actively supported the adoption of the CHASE open science initiative with focus on data sharing for the major Empirical Software Engineering conference so that we could pave the road for a long-term change in that community. Sebastian Uchitel, general chair of the International Software Engineering Conference (ICSE) 2017, further supported an initiative to foster sharing of preprints, and Natalia Juristo, general chair of ICSE 2021, further actively supports the adoption of the broader ESEM open science initiative to our major general software engineering conference. Finally, we want to thank Per Runeson, Klaas-Jan Stol, and Breno de Fran\c{c}a for their elaborate comments on earlier versions on this manuscript.
\end{acknowledgement}

\bibliographystyle{plain}
\bibliography{Literature}
\end{document}